\documentclass[aps,prb,twocolumn,superscriptaddress]{revtex4}
\usepackage{graphicx}
\pdfoutput=1

\begin{document}
\title{Edge states in a honeycomb lattice: effects of  anisotropic hopping and mixed edges}

\author{Hari P. Dahal}
 \affiliation{Theoretical Division, Los Alamos National Laboratory,
Los Alamos, New Maxico 87545}

\author{Zi-Xiang Hu}
\affiliation{Asia Pacific Center for Theoretical Physics, Pohang, Gyeongbuk 790-784, Korea}
\affiliation{National High Magnetic Field Laboratory and
Department of Physics, Florida State University, Tallahassee,
Florida 32310, USA}

\author{N. A. Sinitsyn}
\affiliation{CNLS/CCS-3, Los Alamos National Laboratory, Los Alamos, NM 87545 USA}

\author{Kun Yang}
\affiliation{National High Magnetic Field Laboratory and
Department of Physics, Florida State University, Tallahassee,
Florida 32310, USA}

\author{A. V.  Balatsky}
\affiliation{Theoretical Division, Los Alamos National Laboratory,
Los Alamos, New Maxico 87545}
\affiliation{Center for Integrated
Nanotechnology, Los Alamos National Laboratory, Los Alamos, New
Maxico 87545}\email[] { avb@lanl.gov, http://theory.lanl.gov}
\date{05/28/2009}

\begin{abstract}
 We study the edge states in graphene in the presence of a magnetic field perpendicular to the plane of the lattice. Most of the works done so far discuss the edge states in either zigzag or armchair edge graphene considering an isotropic electron hopping. In practice, graphene can have mixture of armchair and zigzag edges and the electron hopping can be anisotropic, which is the subject of this article. We predict that the mixed edges smear the enhanced local density of states (LDOS) at $E=0$ of the zigzag edge and, on the other hand, the anisotropic hopping gives rise to the enhanced LDOS at E=0 in the armchair edge. The behavior of the LDOS can be studied using scanning tunneling microscopy (STM) experiments. We suggest that care must be taken while interpreting the STM data. It is because the clear distinction between the zigzag edge (enhanced LDOS at $E=0$) and  armchair edge (suppressed LDOS at $E=0$) can be lost if the hopping is not isotropic and if the edges are mixed.
 \end{abstract}

\maketitle
\section{introduction}

Graphene is a one-atom-thick honeycomb lattice of carbon atoms.
 The experimental success of extracting
graphene~\cite{Novoselov1} has attracted multifarious research
activities recently.~\cite{Yzhang,Novoselov,
schedin2007,ohishi2007,
chen2008,wang2008,meyer2007,arikawa2009,castro2009,
geim2007,wakabayashi1999}
 The excitement  in various discipline of physics is originated from graphene's unique two dimensional structure.~\cite{Novoselov}
The tight binding calculations show that the low energy
excitations in graphene are linearly dispersive, hence massless,
and the linear dispersion is confirmed through integer quantum
Hall measurements.~\cite{Novoselov,Yzhang} The Hall conductivity
in these experiments is shown to behave differently from that of
the conventional two dimensional electron system created in
semiconducting hetero structures, namely the conductivity shows
half integer rather than
 the integer effect. The study of the edge states in the presence of the magnetic field
  provides one way of understanding the results of the quantum Hall
  measurements.~\cite{kane2005,butter,abanin2006}

 Several research works have been focused on the study of the edge
states in graphene~\cite{nakada1996,peres2006,Brey, zheng2007,
kohmoto2007,ryu2002, sasaki2006a,
Fujita,sasaki,castro2006,Niimi,Kobayashi,arikawa2009}
 having two types of edges:
 a) armchair edge, and b) zigzag edge.
 In the absence of the magnetic field, in the nearest neighbor hopping approximation, the zigzag edge graphene has non dispersive
 zero energy states at the edge which are also known as surface states.
In the armchair edge graphene these states are absent.  In the
presence of the magnetic field, a bulk graphene has a set of
quantized energy bands (Landau levels),  whereas  both the zigzag and armchair
     edge graphene develop dispersive edge states between the Landau levels.
 The surface states of the zigzag edge graphene survive in the presence of the magnetic field. Therefore the presence (absence)  of the surface states in the zigzag (armchair) edge grahpene signals about the type of the edge a given honeycomb
     lattice has. The presence of the surface states in a zigzag edge graphene gives rise to an enhanced local density of
     states (LDOS) at energy $E=0$ close to the edge.

However, the result that the zigzag edge has the surface states
(characterized by the enhanced local density of states at $E=0$ at
the edge) and the armchair edge does not is
obtained by assuming \textit{isotropic} hopping of the electrons between
 the nearest neighbor carbon atoms. Here, we investigate the effect of possible \textit{anisotropic} hopping on the edge states in graphene.  Whether the hopping can indeed be anisotropic in graphene is  a legitimate question.
      It is easy to conceive  that a graphene lattice on a Si wafer can more likely have anisotropic hopping because of the strain
       induced by the lattice mismatch. Moreover,  one can always apply an
       intentional external strain on the lattice to induce the anisotropic hopping.

If we consider such a possibility of
   the anisotropic hopping of the electrons in graphene, the band structure of the armchair edge graphene changes.
   It has been shown that the zero energy states appear even in the armchair edge
    graphene~\cite{kohmoto2007} at zero magnetic field.
    We numerically determine the LDOS for anisotropic hopping and show that a) the enhanced LDOS at $E=0$ appears in the armchair edge graphene, b)
  in the zigzag edge graphene, the band structure changes slightly but the surface states do not disappear, c)
    the surface states persist in both the zigzag and armchair edge graphene in the presence of the magnetic field.

Most of the studies of the edge states in graphene consider an ideal edge having either only  an  armchair or a zigzag edge.
 It has been shown experimentally that a practical graphene lattice has mixed armchair and zigzag edges.~\cite{geim2007} By mixed edges we refer to a region of the lattice where edges of different geometry cross. We mimic those crossing in a model where a vertical armchair edge crosses a horizontal and a tilted zigzag edge, and a horizontal zigzag edge crosses a tilted zigzag edge.  We study the effects of this edge mixing on the edge states of graphene. The mixture of the different  edges changes  the local density of states at the edges. Here we mainly focus on the effect of the mixed edges on the surface states.
  For our chosen geometry of the edges (see Fig. (\ref{latticefur}))  we show that a) a zigzag edge close to an  armchair edge has reduced LDOS at  $E=0$, b) an  armchair edge close
  to a zigzag edge has enhanced LDOS at $E=0$, c) near the crossing of two zigzag edges the LDOS at E=0 can be completely suppressed.

  The edge states in graphene can be studied experimentally by using a local probe such as scanning tunneling microscopy (STM). The STM experiments measure the differential conductance which is proportional to the density of states. So we have calculated averaged DOS over  a unit hexagonal cell.

The paper is organized as follows. In section II we give an
overview of the edge states in graphene nanoribbon with isotropic
hopping in the absence and presence of the magnetic field. In
section III we discuss the effect of the anisotropic hopping  on
the edge states. The effect of the random mixing of edges is
discussed in section IV. We conclude our work in section V.

\section{Edge states in graphene having isotropic hopping}

The honeycomb lattice of graphene has two nonequivalent lattice
sites (A, B) per unit cell. The Hamiltonian then takes the form of a
matrix. The band structure is calculated using tight binding
model~\cite{wallace,rasito,twodom,ajorio,SReich} where the nearest
neighbor hopping scales the kinetic energy of the electrons. The
diagonalization of the Hamiltonian reveals that the valence and
the conduction bands meet each other at the corners of the
Brillouin zone. The density of states is linear in energy close to
the Fermi energy.

A uniform magnetic field can be included in the tight-binding
Hamiltonian by introducing the Peierls phase characterized by the
inclusion of a magnetic vector potential $\vec{A}$ in the electron
hopping term
\begin{equation}
H = -\sum_{ij} t_{ij} c_j^+ c_i e^{i2\pi \phi_{ij}}
\label{tbh}
\end{equation}
where i and j are the nearest-neighbor sites, $t_{ij}$ is the nearest neighbor hopping energy, $c_i$ and $c_i^+$ are fermion annihilation and creation
operators respectively, and $\phi_{ij}$ is the phase factor which is given by the line integral of the magnetic vector potential as:
\begin{equation}
\phi_{ij} = \frac{e}{hc} \int_i^j \vec{A} \cdot d\vec{l}.
\label{phase}
\end{equation}

\begin{figure}
\includegraphics[width=8cm, height=5.5cm,angle=0]{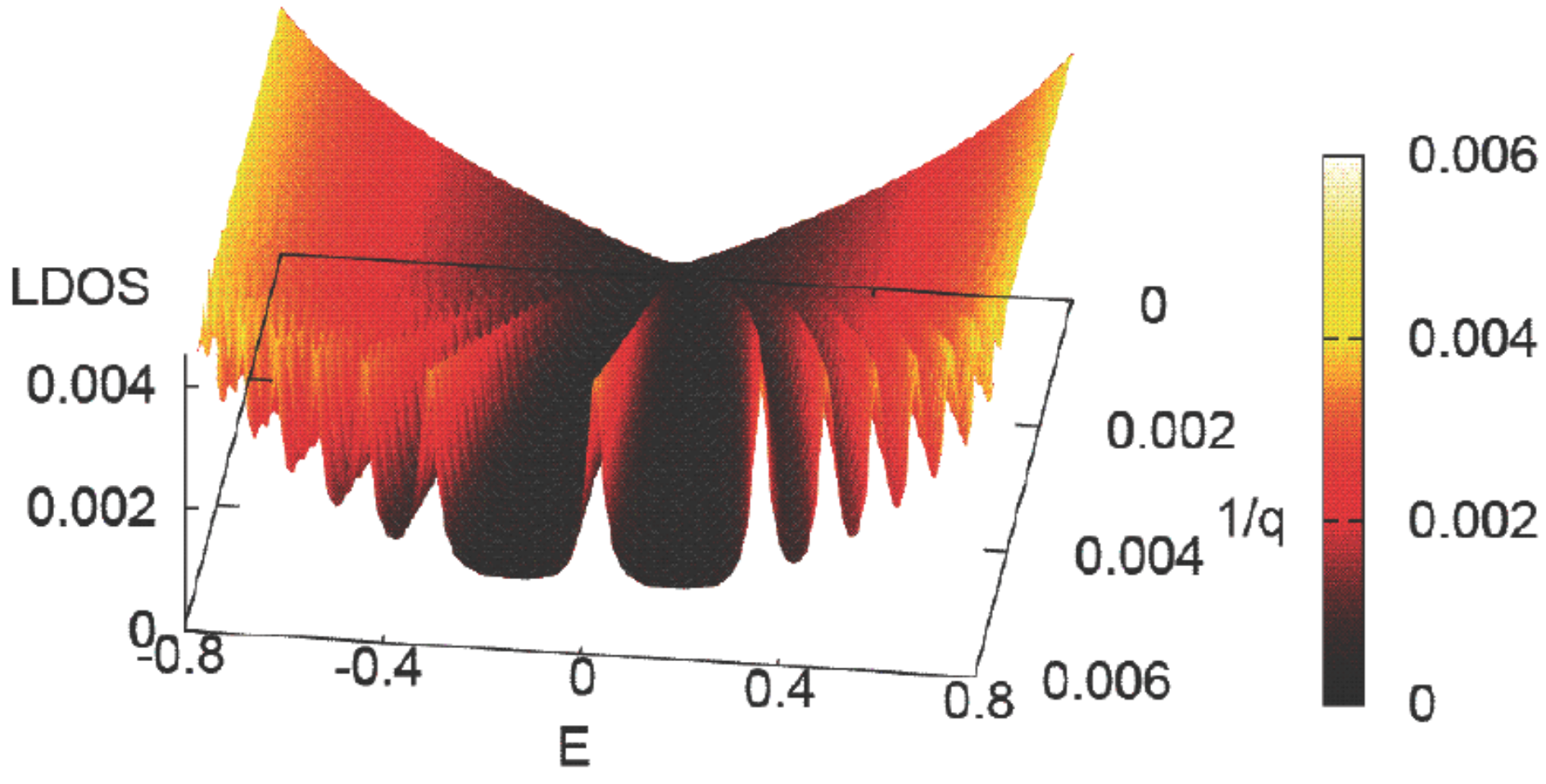}
 \caption{(color online) The density of states as a function of magnetic field and energy is shown.
 The Landau levels are clearly resolved as indicated by the enhanced density of states.
 The small decay of the amplitude of the density of states at higher energy is due to the finite size effect.
 The magnetic field is expressed in terms of the magnetic flux quantum ($\phi=\frac{BA}{2\pi}=\phi_0/q$). The energy scale is the hopping
 term t which is set to unity.}
 \label{energy_dos_mag}
\end{figure}

 We write the Hamiltonian matrix for different geometries of the graphene lattices such as bulk, armchair edge, zigzag edge and mixed edge graphene. We diagonalize the Hamiltonian to get the eigen-values and the eigen-vectors which is used to calculate the local density of states. First, we reproduce the results obtained based on the standard assumption of the isotropic hopping. Many of these results are discussed in several other literatures. ~\cite{ Brey, zheng2007,peres2006,
kohmoto2007,ryu2002, sasaki2006a,
Fujita,sasaki,castro2006,Niimi,Kobayashi} We have presented them here to facilitate the discussion of the results obtained for the anisotropic hopping and the mixed edges. For bulk graphene we reproduce the linear in $\textbf{k}$ dispersion of the quasi-particle excitations and the linearly vanishing density of states at the Fermi energy. In the presence of the magnetic field eigen states  split into discrete bands
(Landau Levels). No state is allowed between the Landau levels without disorder.
  The density of states for this system is shown in
   Fig. (\ref{energy_dos_mag}). From this result the relation between the Landau level (LL)
energy and Landau level index $n$ can be extracted.
 It can be shown that the energy is proportional to the square root of the LL index,
 and  the magnetic field.~\cite{kun} These Landau levels have been observed in STM experiments.~\cite{li2008}

\begin{figure}
\includegraphics[width=8cm,height=5cm]{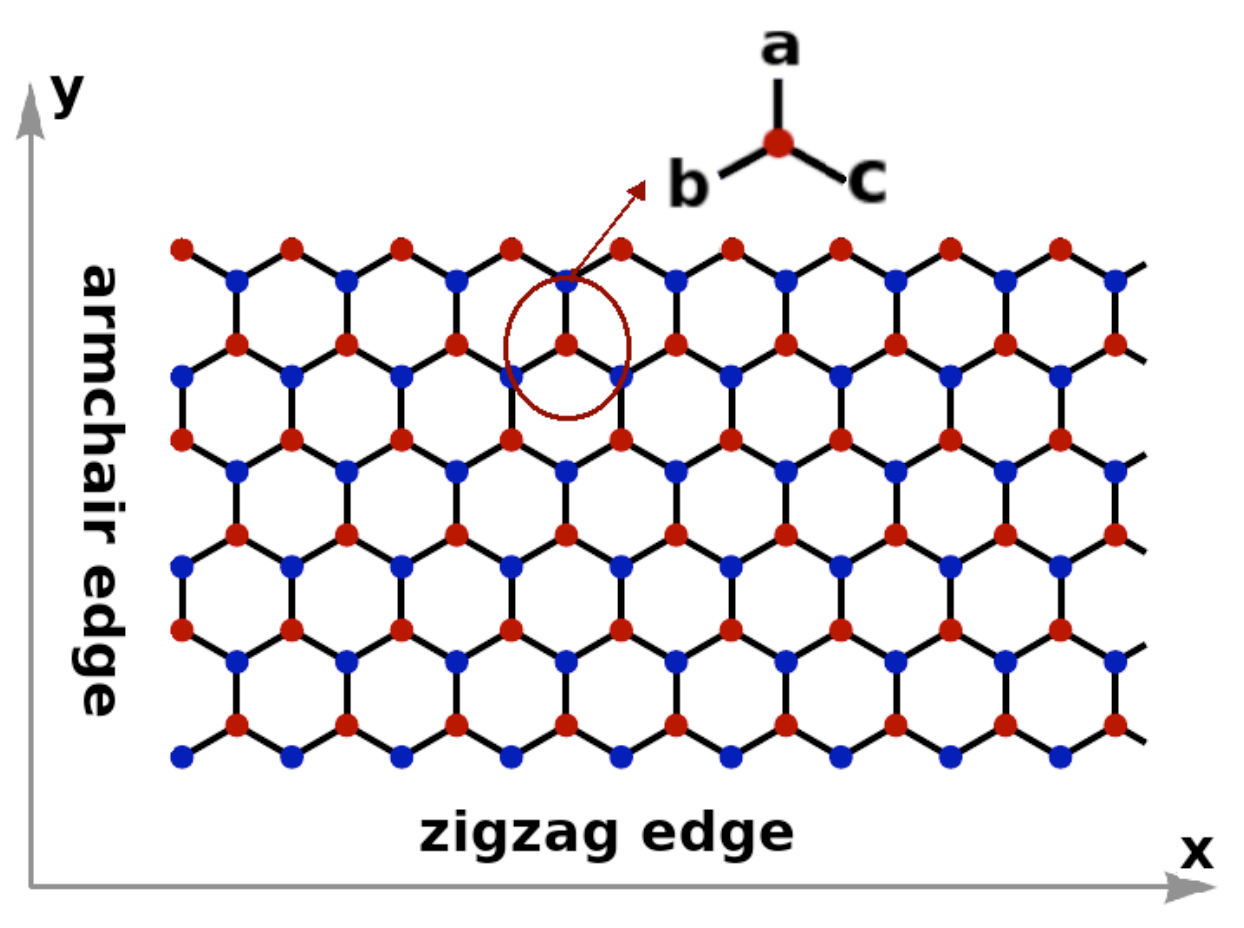}
\caption{\label{lattice}(color online) The lattice structure of
the graphene sheet with zigzag edge and armchair edge. While
considering the armchair edge, we rotate it by $90^0$ to make the
edge along X-axis. The directions of the bonds are labelled by a,
b and c.}
\end{figure}

\begin{figure}
 \includegraphics[width=5cm, height=8.0cm,angle=-90]{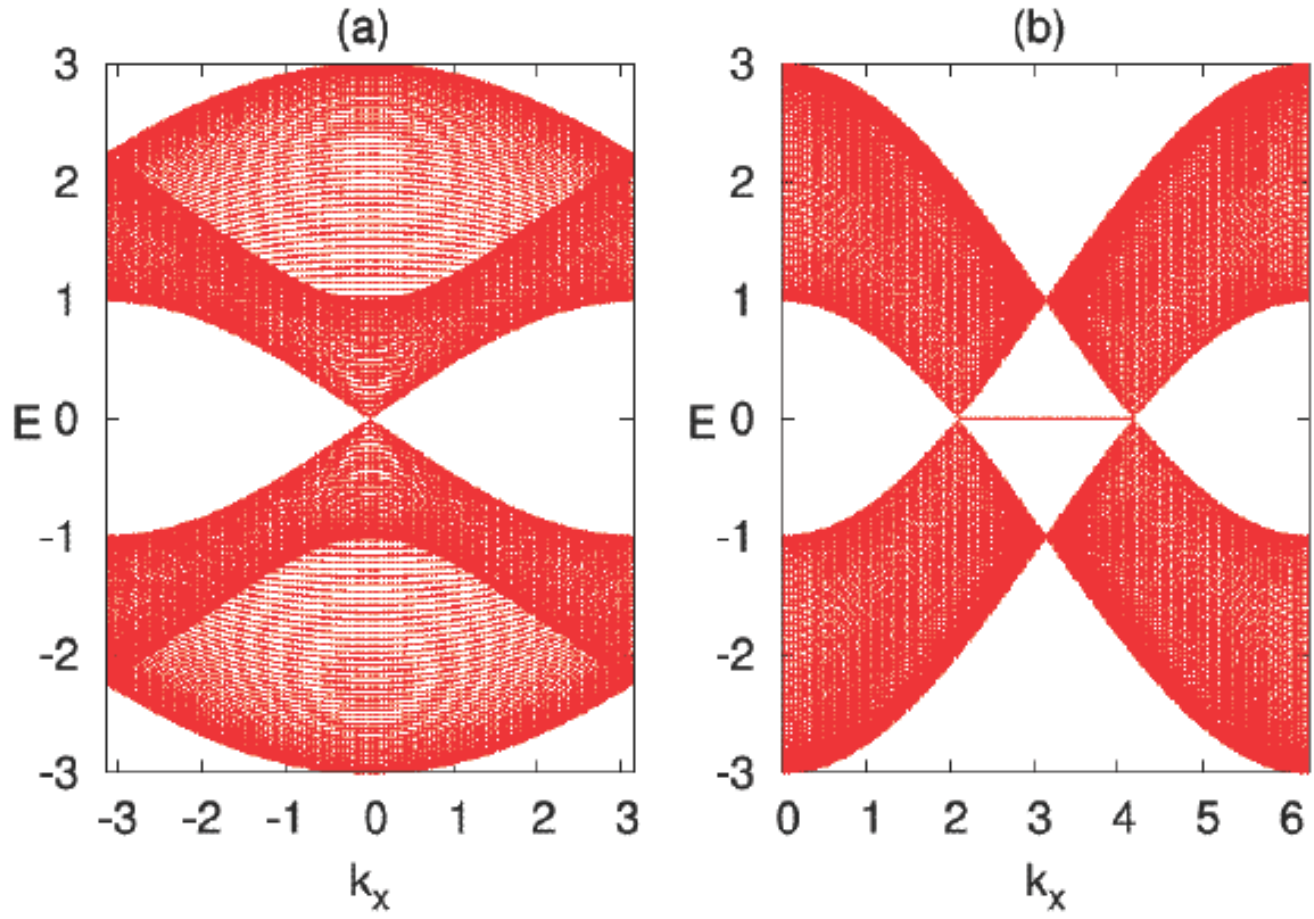}
 \caption{(color online) The dispersion relation of the electrons in a) armchair and b) zigzag edge graphene in the absence of the magnetic field is shown.
 The zigzag edge graphene has zero energy states whereas they are absent in the armchair edge graphene.
 The zero energy states correspond to the localized surface states of the zigzag edge graphene.}
 \label{energy_nomag}
\end{figure}

 For the study of the edge states we assume either one of the edge along the X-axis (Fig. \ref{lattice}) and write a tight-binding Hamiltonian for a lattice
 having 600 lattice sites along X-axis and a periodicity along Y-axis with 400 repetition.
For zigzag edge graphene only one type of  sublattice, say A, is at the edge. In armchair edge graphene
 both A and B sublattices are at the edge.
 The dispersion relation of the electrons in the armchair and zigzag edge graphene in the absence of the magnetic
 field are shown in Fig. (\ref{energy_nomag}).

The dispersion relation of the electrons in the armchair edge graphene is
qualitatively similar to that of the bulk graphene, i.e., the
dispersion is linear in $\textbf{k}$ close to the half filling point. The valence and
 conduction bands meet at the Fermi point,  depending upon the width of the nano
ribbon a gap can open up in the  armchair edge graphene.~\cite{zheng2007} In the case of
zigzag edge graphene the dispersion relation is quite
different from that of the armchair edge and bulk graphene due to the presence of
the zero energy states.\cite{peres2006, nakada1996} The zero energy states of the zigzag edge graphene are called the surface states.

The dispersion relation of the electrons in the zigzag and armchair edge graphene has been obtained by solving the Dirac equation with proper 
boundary conditions. \cite{Brey} The boundary condition for the zigzag edge is such that the electron wave function vanishes on a single sublattice which is on the edge. It gives rise to particle and hole like bands along with evanescent wave functions localized at the surface. These localized states become the zero energy surface states. For the armchair edge graphene the wave function should vanish on both the sublattices at the edges leading to the mixing of the two valley states. These boundary conditions rightly lead to surface states (no surface states) in the zigzag (armchair) edge graphene.

 We calculate the local density of states (LDOS) close to and away from the edge.
   \begin{figure}
 \includegraphics[width=5.0cm, height=8.5cm,angle=-90]{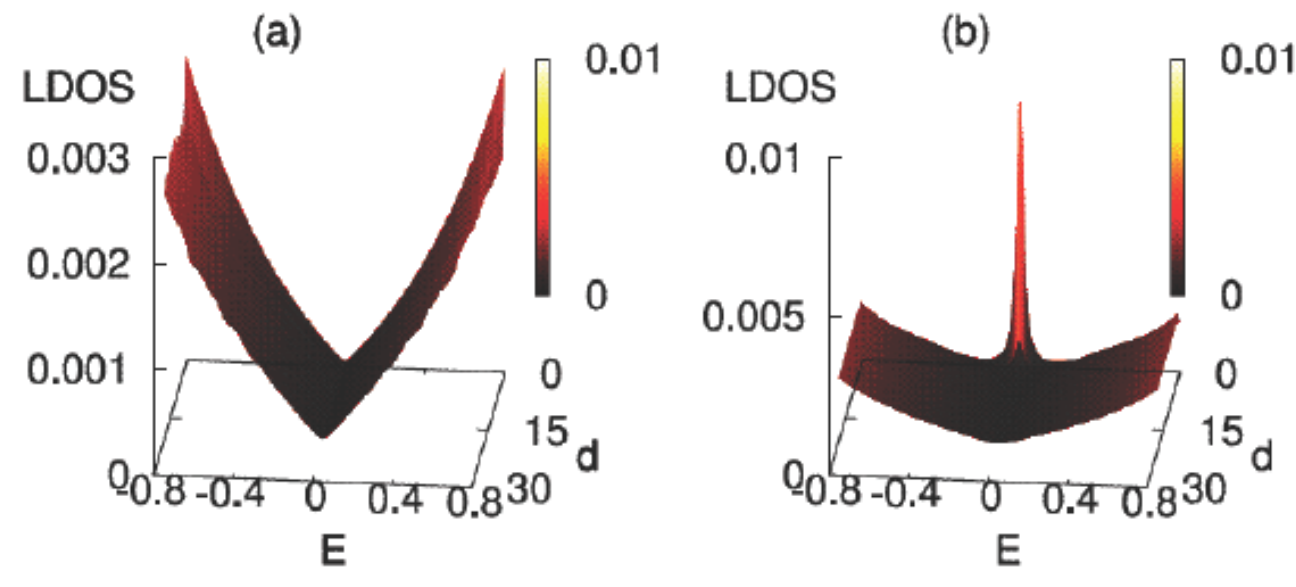}
\caption{(color online) The LDOS as a function of the distance
 from the edge in the absence of the magnetic field is shown for a) armchair edge and
  b) zigzag edge. The LDOS is the averaged LDOS over each hexagon.
  The label axis $d$ represents the position of the hexagon away from the edge.
  The enhanced LDOS close to the zigzag edge graphene signifies the
  surface states. Such states are absent in the armchair edge.}
\label{ldos_nomag}
\end{figure}
The LDOS at the sublattices  A and B close to the zigzag edge is
not the same. The sublattice A    has
enhanced   LDOS at $E=0$ whereas the sublattice B has zero LDOS at $E=0$. In the armchair
edge graphene the LDOS at A and B are equal and it is zero
at $E=0$. We average the local density of states over the six lattice sites of the hexagon, and study its variation as a function of distance of each hexagon away from the edge. The result is shown in Fig. (\ref{ldos_nomag}).
 In this figure ''d" represents the number
of the hexagonal cell away from the edge. We see in this figure that there is an
enhanced LDOS at $E=0$  at the zigzag edge
 but there is  no weight of the LDOS at  $E=0$ in the armchair edge. The amplitude of the density of state peak decays sharply away from the zigzag edge.

\begin{figure}
 \includegraphics[width=5cm, height=8.50cm,angle=-90]{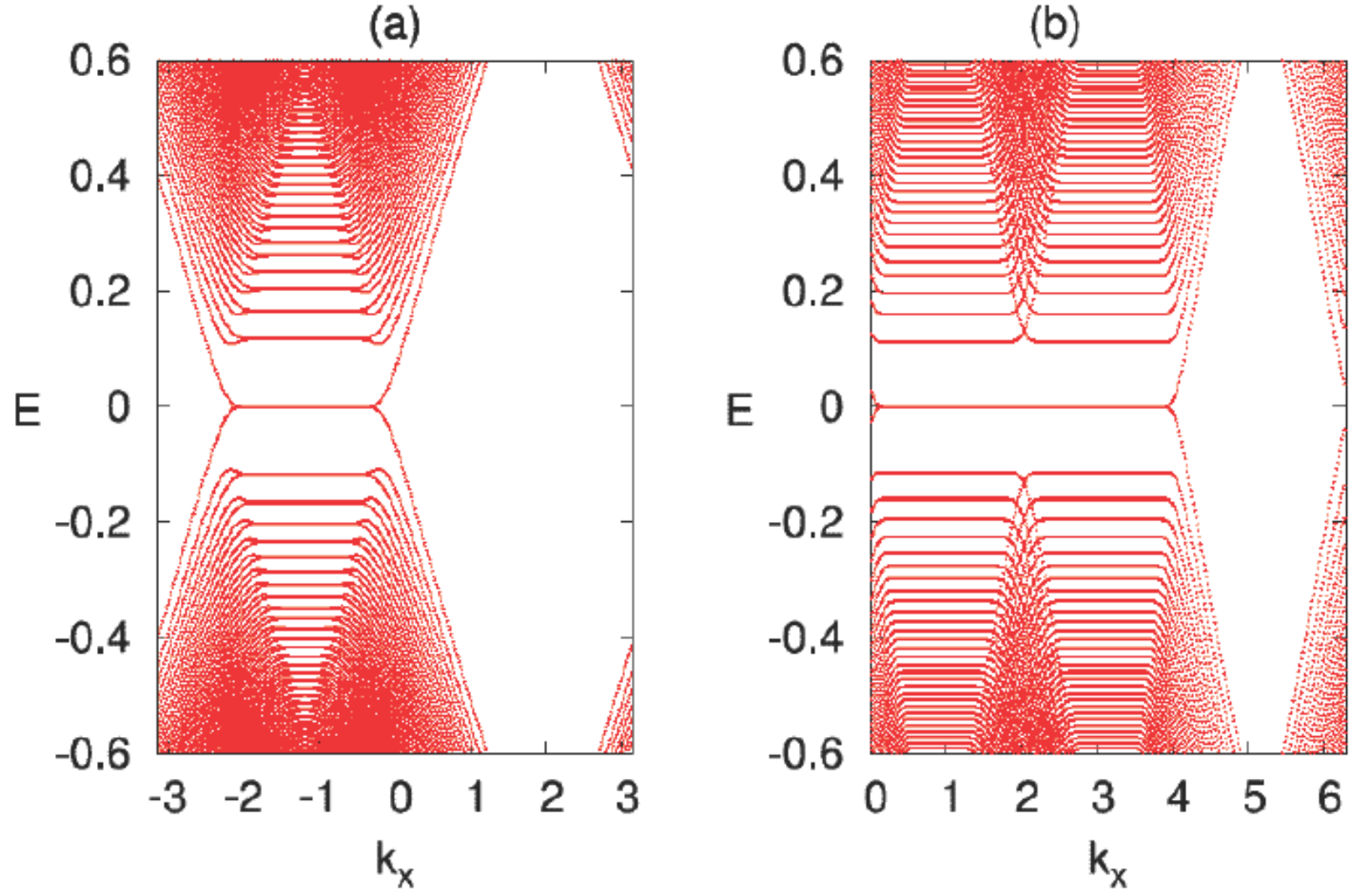}
 \caption{ The dispersion relation of the electrons in a) armchair and b) zigzag edge graphene in the presence
 of the magnetic field. The magnetic field is such that $\phi=\frac{\phi_0}{825}$. }
 \label{energy_mag}
\end{figure}

 We repeat the above discussed comparison of the edge states when the magnetic field is applied perpendicular to the graphene lattice.  Fig. (\ref{energy_mag})
  shows the dispersion relations of the electrons in the armchair and zigzag edge graphene.
We see the emergence of additional states between the quantized Landau level. These are the edge states. The edge states are gapless and dispersive. There is  a significant difference  between the edge states of the armchair and
zigzag edge graphene.\cite{peres2006,Brey}  Although the number of edge states
branches below certain energy is equal for both types of edges, they originate in pairs in the armchair edge graphene (except for the lowest Landau level). We would like to note here that these dispersion relations have been obtained by solving the Dirac equation with appropriate boundary conditions and using the tight binding calculation. \cite{Brey, abanin2006, peres2006,arikawa2009,castro2006}   In the presence of the magnetic field there are zero energy states in both the armchair and zigzag edge graphene. To distinguish between the n=0 Landau level states (which are zero energy states) and the surface states we need to calculate the local density of states.

\begin{figure}
 \includegraphics[width=8.5cm, height=5.0cm]{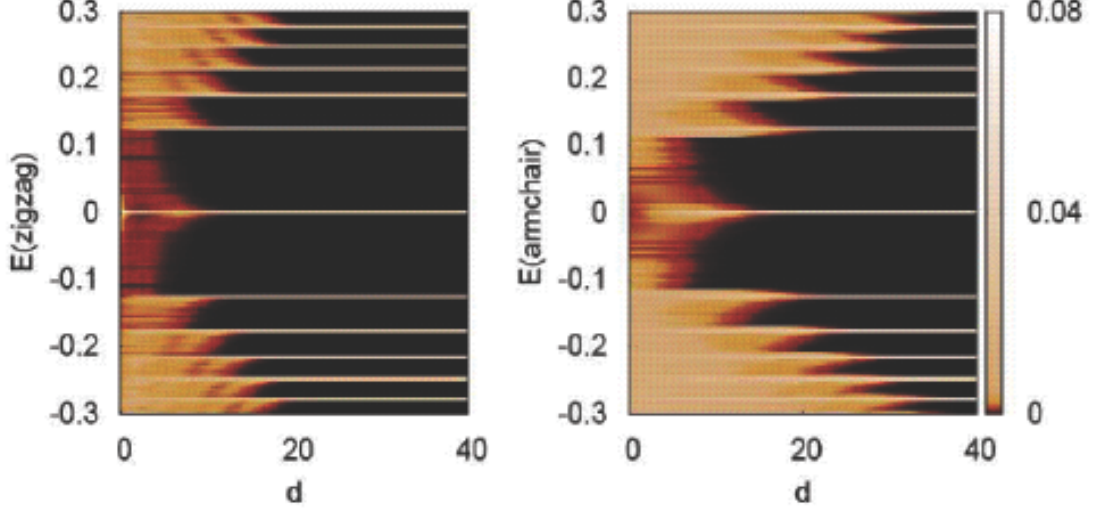}
 \caption{(color online) The 2D plot of LDOS as a function of the distance of the hexagonal cell from the edge for a) zigzag edge and b) armchair edge.
 d represents the number of the hexagonal cells away from the edge.
 This figure clearly shows the surface states and the dispersive edge states in the zigzag edge graphene.
 It also shows the edge states in armchair edge graphene. The magnetic field is quantified by the flux per unit cell: $\phi=\phi_0/701$.}
 \label{ldos_mag}
\end{figure}

 We calculate the LDOS on the hexagonal
 cells as a function of their distance (d) from the edge. The results are shown in Fig. (\ref{ldos_mag}).
 In the left figure we can see that there is an enhanced LDOS at $E=0$, which corresponds to the surface states of the zigzag edge.
 These surface states are absent near the armchair edge. On the other hand, in the presence of the
 magnetic field, the zigzag edge and the armchair edge give rise to
 finite LDOS  between the Landau levels close to the edges.
 It characterizes the dispersive edge states.
  Similar results have been presented in Ref. [\onlinecite{abanin}] by Abanin et. al., using effective Dirac Hamiltonian near the Dirac point.
   In contrary our results rely on the full tight-binding Hamiltonian. Note that we can see the dispersive edge states more clearly.

Next, we consider the possibility of the anisotropic hopping of the electrons between the nearest neighbor carbon atoms. We study the
effect of this anisotropy on the edge states of the armchair and zigzag edge graphene in the
absence and presence of the magnetic field.

\section{Edge states in graphene having anisotropic hopping}

We use the similar geometry as shown in Fig. (\ref{lattice}) to study the
effects of the anisotropic hopping on the edge states of graphene. The directions of the
bonds in graphene are labelled by a, b and c.
 In the case of the zigzag edge, by changing the hopping energy along any combination
of the three bonds, we did not see any qualitative change in the band
structure, especially the surface states. In the absence of
the magnetic field, there are zero energy states which give rise to the enhanced LDOS at $E=0$ at the
edge. In the presence of the magnetic field the dispersive edge
states appear and the enhanced LDOS at $E=0$ corresponding to the
surface states are also present.

\begin{figure}
 \includegraphics[width=8.0cm, height=5.0cm, angle=0]{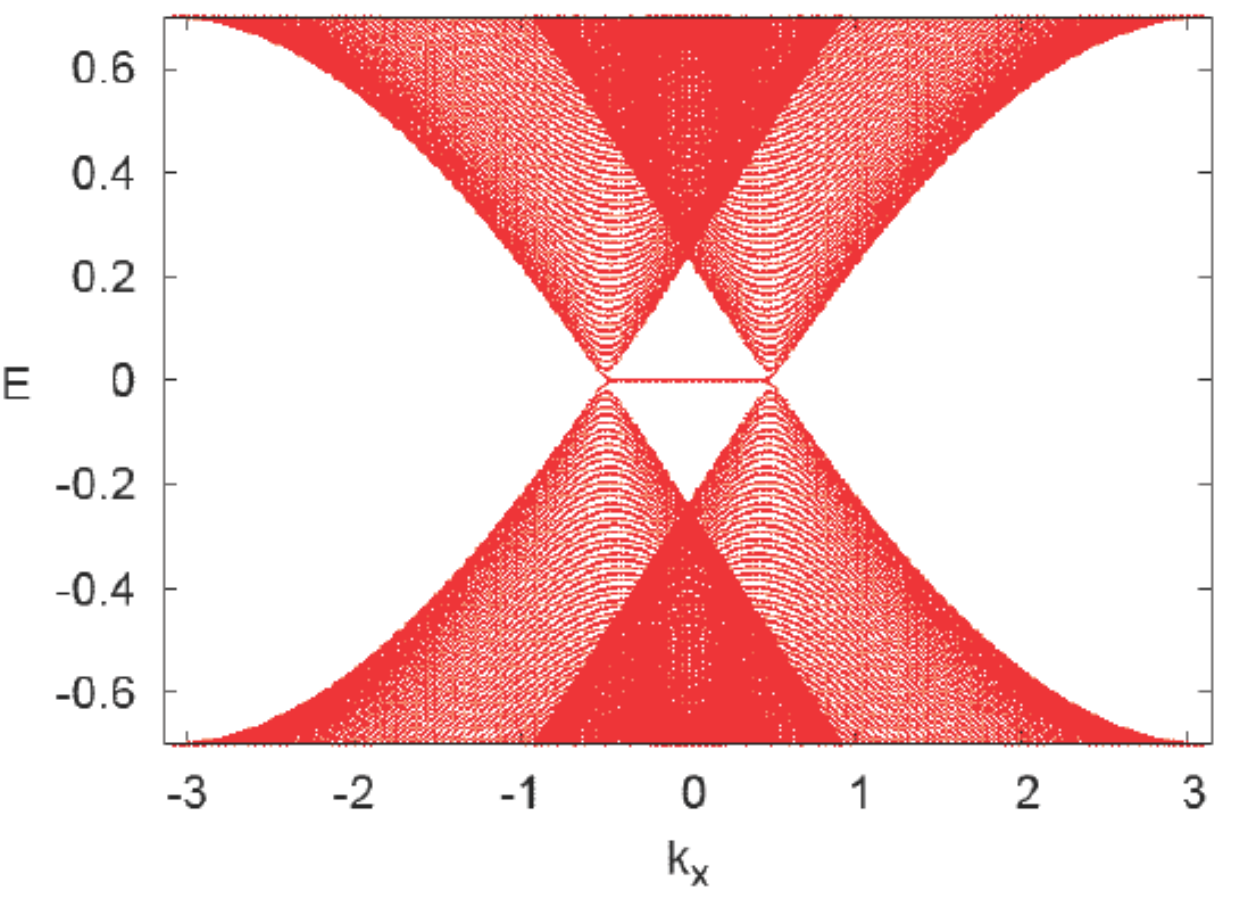}
 \caption{The dispersion relation of the armchair edge graphene having
anisotropic hopping along the bonds in the direction b or c in the
absence of magnetic field. The hopping energy is reduced by $30\%$
compare with the unchanged bonds which have unity hopping. We can
see the presence of the zero energy states similar to that of the
isotropic and anisotropic zigzag edge graphene. }
 \label{energy_assym_nomag}
\end{figure}

\begin{figure}
  \includegraphics[width=8.5cm, height=5.0cm,angle=0]{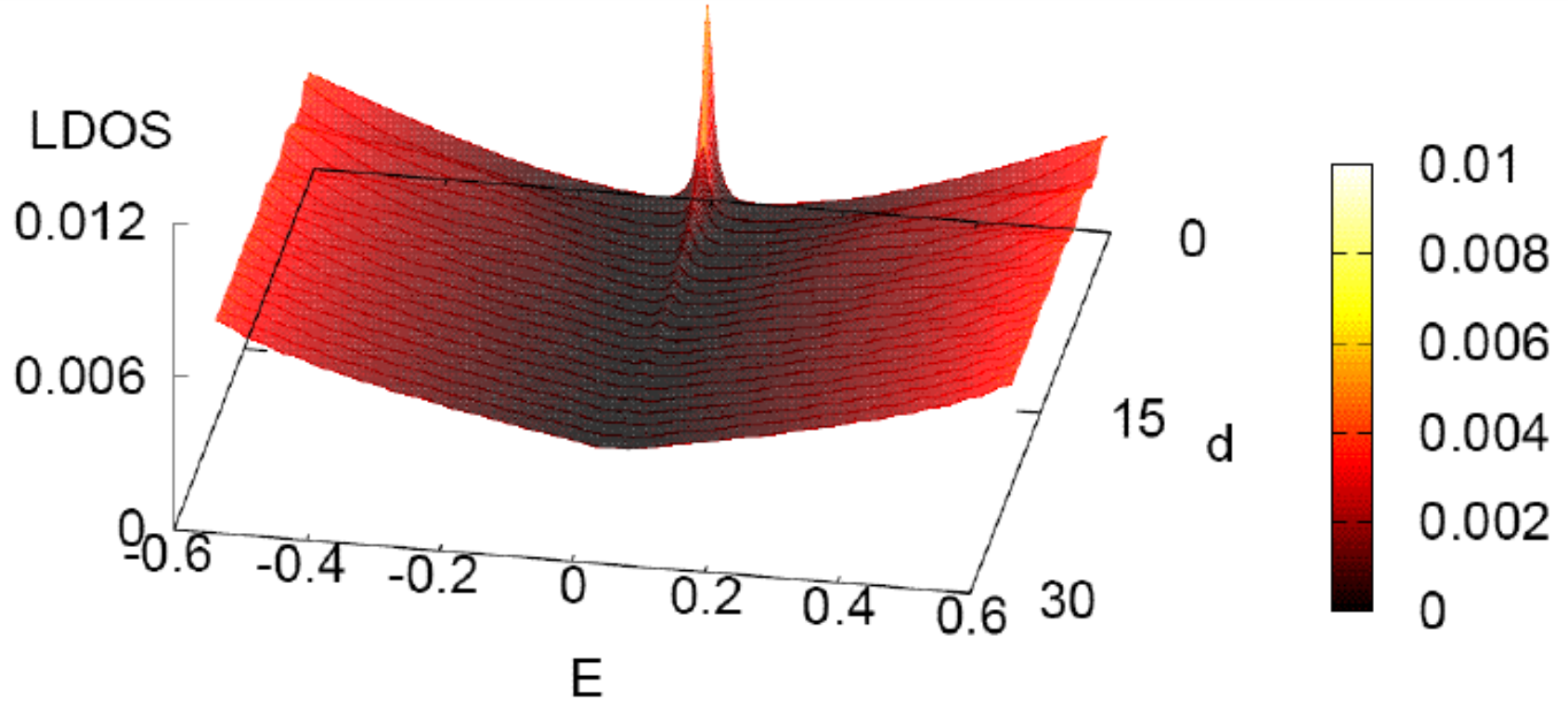}
 \caption{(color online) The LDOS as a function of the distance of a hexagonal cell away from the edge
  of the armchair edge graphene having anisotropic hopping in the absence of magnetic field.
  We see an enhanced LDOS at $E=0$ which corresponds to non dispersive the surface states.
  This behavior is similar to that of the isotropic zigzag edge graphene.}
 \label{ldos_assym_nomag}
\end{figure}

The situation is different in the armchair edge graphene. When the
armchair edge is along the X-axis and the ribbon repeats along the
Y-axis (Fig.(\ref{lattice}) rotated by $90^0$), let us
assume that the hopping energy along direction ``a" is reduced. In
this situation the
dispersion relation remains similar to that of the isotropic
case (note that we are focusing our discussion on the edge states).
But the band structure change dramatically when the hopping along the
angled bond ''b" or ``c" is changed. (If we change the hopping
along both the angled bonds with equal magnitude, the band
structure again becomes qualitatively the same to that of the
isotropic armchair edge graphene.) When the hopping along  the ''b'' or ''c'' bonds is changed, the band structure of the
armchair edge graphene becomes qualitatively similar to that of the
zigzag edge graphene. The presence of the zero energy
states as shown in Fig. (\ref{energy_assym_nomag}) (compare with Fig. (\ref{energy_nomag}b)) is one of the examples. Now the electron wave
function also changes, namely it is localized at the edge of the
lattice.
Fig. (\ref{ldos_assym_nomag}) shows the LDOS as a function of the
distance of the hexagonal cell away from the edge. There is a clear enhancement in the LDOS at $E=0$; which is the
generic behavior of the zigzag edge graphene.

\begin{figure}
 \includegraphics[width=8.5cm, height=5.0cm, angle=0]{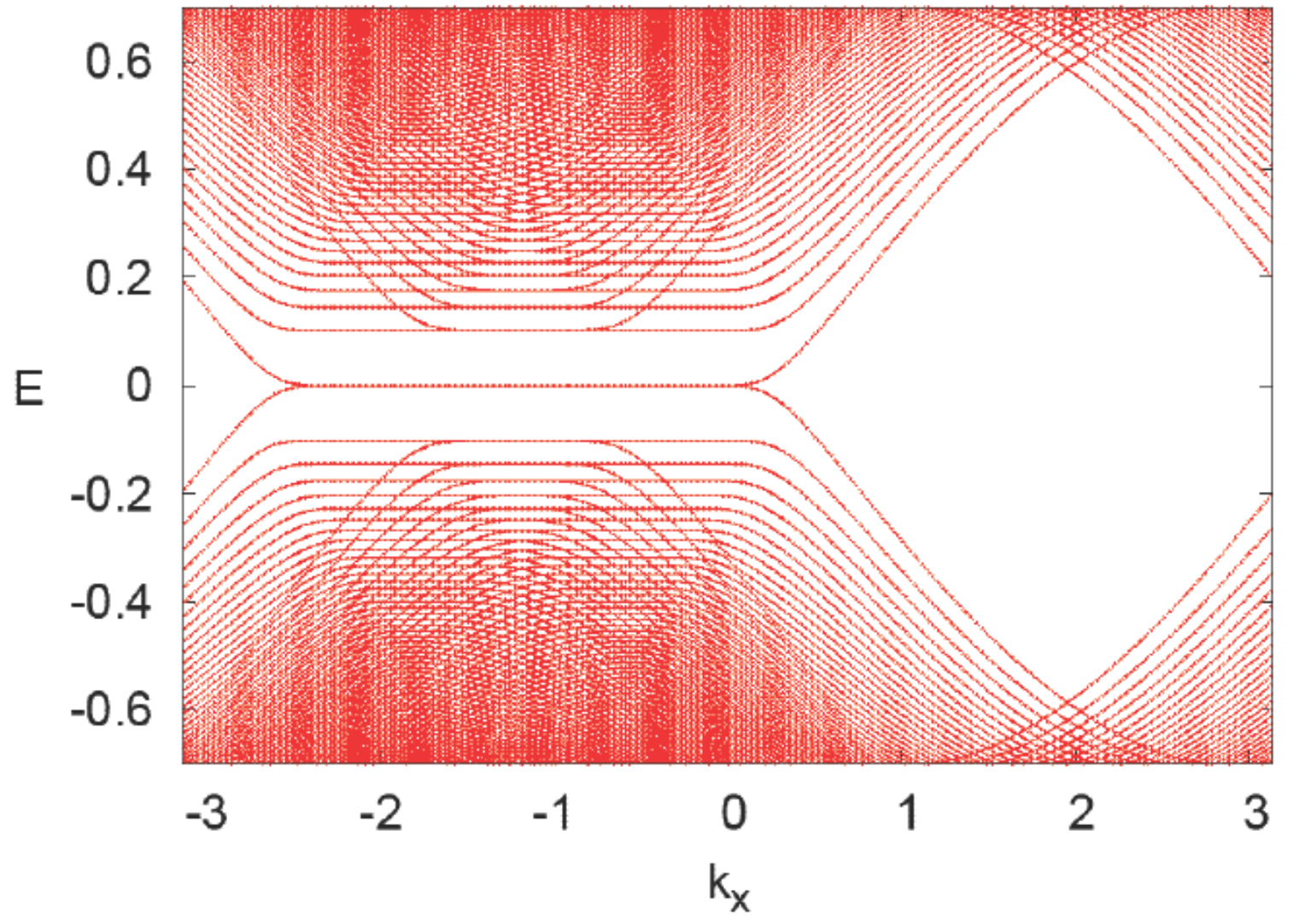}
 \caption{ The dispersion relation of the armchair edge graphene having anisotropic
 hopping along the bonds ``b" or ``c" (Fig.\ref{lattice}) in magnetic field.
  It is similar to that of the isotropic zigzag edge graphene in the presence of
   the magnetic field. The similarity is inferred from the disappearance of the pairwise dispersive edge
    states and appearance of the unpaired edge states as it is seen in the zigzag edge graphene. The
    magnetic field is given by $\phi=\frac{\phi_0}{825}$.}
 \label{energy_assym_mag}
\end{figure}

\begin{figure}
  \includegraphics[width=8.5cm, height=5.0cm, angle=0]{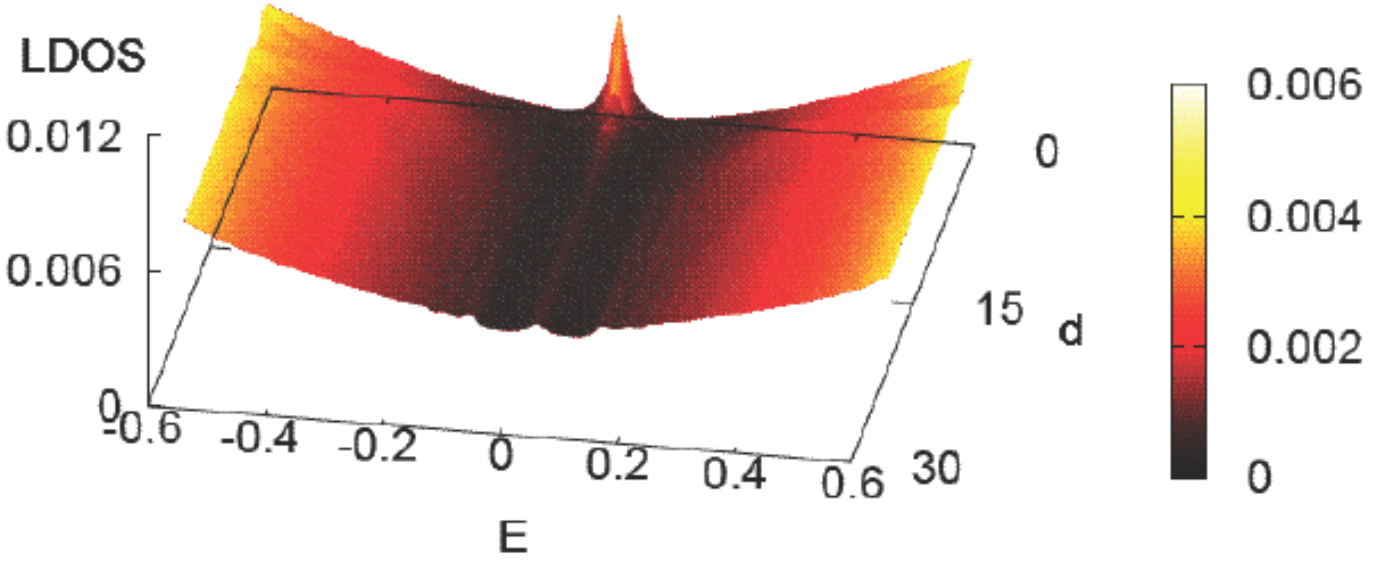}
 \caption{(color online) The enhancement of the LDOS at $E=0$ is shown in the armchair edge graphene having anisotropic
  hopping along the bonds b or c in a magnetic field given by $\phi=\frac{\phi_0}{825}$.
  The LDOS behaves similar to that of the zigzag edge graphene.}
 \label{ldos_assym_mag}
\end{figure}

In the presence of the magnetic field the dispersion relation of the electron in
 the armchair edge graphene (having anisotropic hopping along the
bond ''b'' or ''c'') looks similar to that of the zigzag edge graphene as
shown in Fig. (\ref{energy_assym_mag}) (compare with Fig. (\ref{energy_mag})b). The main similarity is
that the edge states now do not come in pair (see Fig. \ref{energy_mag}a for isotropic case) which is similar to what is seen in the zigzag edge graphene (see Fig. \ref{energy_mag}b).
The wave function also behaves similar to that of the zigzag edge
graphene in the presence of the magnetic field. The wave functions
corresponding to the surface state and the edge states are
localized at the edge.  The LDOS is also similar to that of the  zigzag edge
graphene as shown in Fig. (\ref{ldos_assym_mag}). There is an
enhanced LDOS at zero energy at the edge.

The effect of the anisotropic hopping on the edge states of the armchair edge graphene  can be understood by using the concept of induced guage field due to the lattice deformation developed by Sasaki et al. ~\cite{sasaki2006} Let's denote the modulation of the hopping along a, b, c bonds 
(see Fig. (\ref{lattice}) ) by $\delta \gamma_1$,  $\delta \gamma_2$, and  $\delta \gamma_3$  respectively. For small modulation of the hopping ($\delta \gamma_i <t_{ij}$) the induced gauge field can be written as, 
$v_F A_x(\textbf{r})=\delta \gamma_1(\textbf{r}) - \frac{1}{2}(\delta \gamma_2(\textbf{r}) + \delta \gamma_3(\textbf{r}))$, and $v_F A_y(\textbf{r})=\frac{\sqrt{3}}{2} (\delta \gamma_2(\textbf{r}) - \delta \gamma_3(\textbf{r}))$,  where $v_F$ is the Fermi velocity and 
$\textbf{A}(\textbf{r})=(A_x(\textbf{r}),A_y(\textbf{r}))$ is the deformation induced vector potential.  The corresponding "effective magnetic field" is given by, $B_z(\textbf{r})=\frac{\partial A_y(\textbf{r})}{\partial x}-\frac{\partial A_x(\textbf{r})}{\partial y}$. A finite effective field due to the modulation of the hopping results into the zero energy edge sates. 

Let's apply these relations in the strained armchair edge graphene lattice. In our calculation we have taken position independent $\delta \gamma_1,\delta \gamma_2,\delta \gamma_3$. Then away from the  armchair edge the effective magnetic field is zero because the vector potential is uniform.  Along the edge the situation is different; we can show that the vector potential is nonuniform. Since the armchair edge is taken along the X-axis  $A_x$ is independent of $y$ along the edge. But, $A_y(x)$ changes along the edge between two neighboring sublattices. Lets take two sublattices at (0,0) and (x,0) at the edge. Then one possibility is that $v_F A_y(0,0)=\frac{\sqrt{3}}{2}\delta \gamma_2$ and $v_F A_y(x,0)=\frac{\sqrt{3}}{2}\delta \gamma_3$  (see Fig. 2 interchanging the X and Y axes ). So, $ A_y(x,0) - A_y(0,0)  \sim    | \delta \gamma_2 - \delta \gamma_3 | $ which, for $ \delta \gamma_2 \neq \delta \gamma_3$, results into a nonzero local effective magnetic field. Now it is clear that the anisotropic hopping combined with one missing carbon-carbon bond per carbon atom at the edge give rise to zero energy states.  

Analytical calculations show that the lattice-deformation induced effective guage field shifts the position of the Dirac cones in the momentum space. ~\cite{manes2007, guinea2008}  The shift of the Dirac cones at K and K'  is opposite to each other and proportional to the induced vector potential. The result presented in  Fig.  \ref{energy_assym_nomag} is consistent with the result of the analytical calculations.

From the above discussions we conclude that the strained armchair edge
graphene can have similar edgestates as that of the zigzag edge
graphene.  We want to point out that in the armchair
edge graphene the amplitude of LDOS at $E=0$ depends upon the
strength of the anisotropy, the smaller the anisotropy the smaller is the
amplitude. Next we calculate the LDOS in the vicinity of the edge crossing to study the effect of the edge mixing on the
edge states of graphene.

\section{mixed edge graphene}

\begin{figure}
 \includegraphics[width=8.5cm, height=5.5cm]{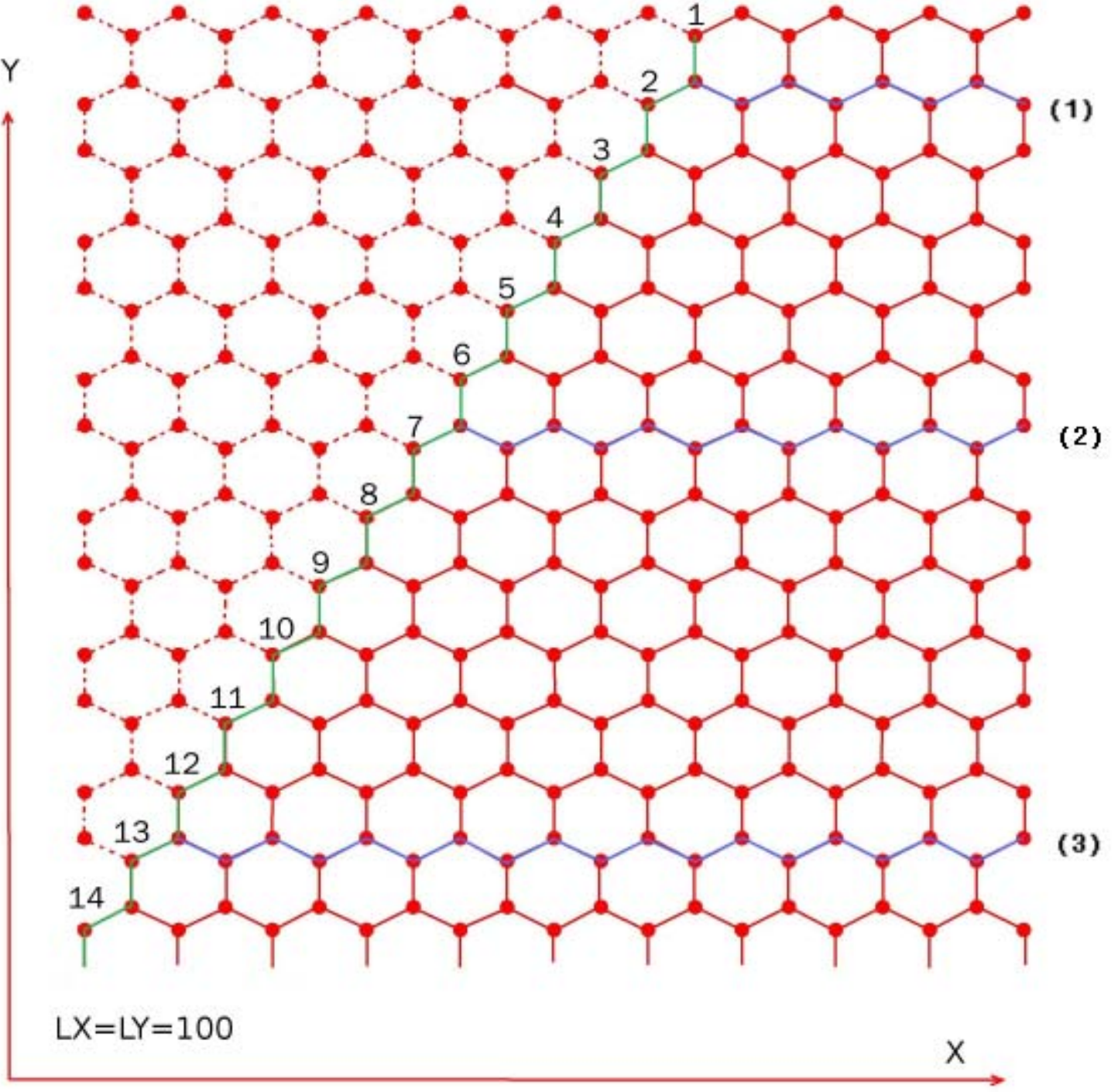}
 \caption{\label{latticefur}(color online)  The schematic figure of the arbitrary edge graphene.
 The length $LX=100$ and the width $LY=100$ means there are 50 hexagonal cells in each direction.  There are two zigzag edges parallel to the X-axis,
 two armchair edges parallel to the Y-axis and one tilted zigzag edge (guided by a green line).
 The dotted lines represent the part of the lattice where there are missing carbon atoms.
 }
 \label{lattice_fur}
\end{figure}

To study the effects of the mixed edge on the edge states of graphene, instead of
considering the graphene ribbon, we directly diagonalize a finite
size system with length $LX$ and width $LY$ as shown in Fig.
(\ref{lattice_fur}).  The size of the lattice is given by $LX=100$ and $LY=100$ with open boundary conditions which means that the sizes in
both of the directions have the length of 50 hexagonal cells.
 Theoretically, there are many ways to form a lattice with mixed edges. The one we choose actually represents a lattice where
 there is a tilted zigzag edge along the green line in in Fig.
(\ref{lattice_fur}), two vertical armchair edges   and
two horizontal zigzag  edges. We diagonalize the tight-binding Hamiltonian for the whole lattice, and calculate the LDOS for each lattice site.  As mentioned above, in the discussion of the zigzag edge,
  we denote the lattice site at the edge by sublattice A.

We first imagine the lattice without the tilted zigzag edge in the presence of the magnetic field. Our calculation shows that
all the hexagonal cells along the horizontal zigzag edge have enhanced LDOS at $E=0$
(we would like to remind here that we calculated the LDOS averaged over the six sites of the hexagon).
Along a line parallel to the Y-axis from the middle of the zigzag edge
the LDOS at E=0  first decreases sharply and then increases gradually to give the bulk value deep inside the lattice.
Along the armchair edge,  all the hexagonal cells have suppressed LDOS at
  $E=0$ except when theses cells are close to the zigzag edge where due to the proximity effect these cells acquire finite LDOS at E=0. Along a line parallel to  the X-axis from the middle of the armchair edge,
the LDOS at E=0 increases gradually from zero to give the bulk value deep inside the lattice.

 Next, we consider the tilted zigzag edge which is introduced as shown in Fig. (\ref{lattice_fur}). The figure shows that the tilted zigzag edge crosses one horizontal zigzag edge and one vertical armchair edge.
Due to the presence of the two zigzag edges of different geometry, for the technicality of the discussion, we
need to redefine the type of atoms which are sitting at the edges of the horizontal and tilted zigzag
edges. We can see from the figure that if the edge atom on the horizontal zigzag edge is denoted by sublattice A, the edge atom on the tilted edge will be sublattice B. Lets redefine the sublattice of the tilted zigzag edge by A'(=B with respect to the horizontal zigzag edge). Then close to these two edges, any sublattice of type A (B) corresponding to the horizontal zigzag edge will be of type B' (A') for the tilted zigzag edge. Now, since a sublattice A (B) or  A' (B')  always has enhanced (suppressed) LDOS at E=0, what will be the resultant LDOS on a hexagonal cell that has sites which simultaneously behave like sublattice A (B') and B (A') for the horizontal (tilted) zigzag edge? Similar question can be asked for the hexagonal cell which are close to the crossing of the vertical armchair and tilted zigzag edges. In the latter case a sublattice A close to the zigzag edge (which has enhanced LDOS at E=0) is also a sublattice A for the armchair edge (which has zero LDOS at E=0).  In what follows we have calculate the LDOS to address these scenarios.

 \begin{figure}
  \includegraphics[width=8.5cm, height=5.0cm]{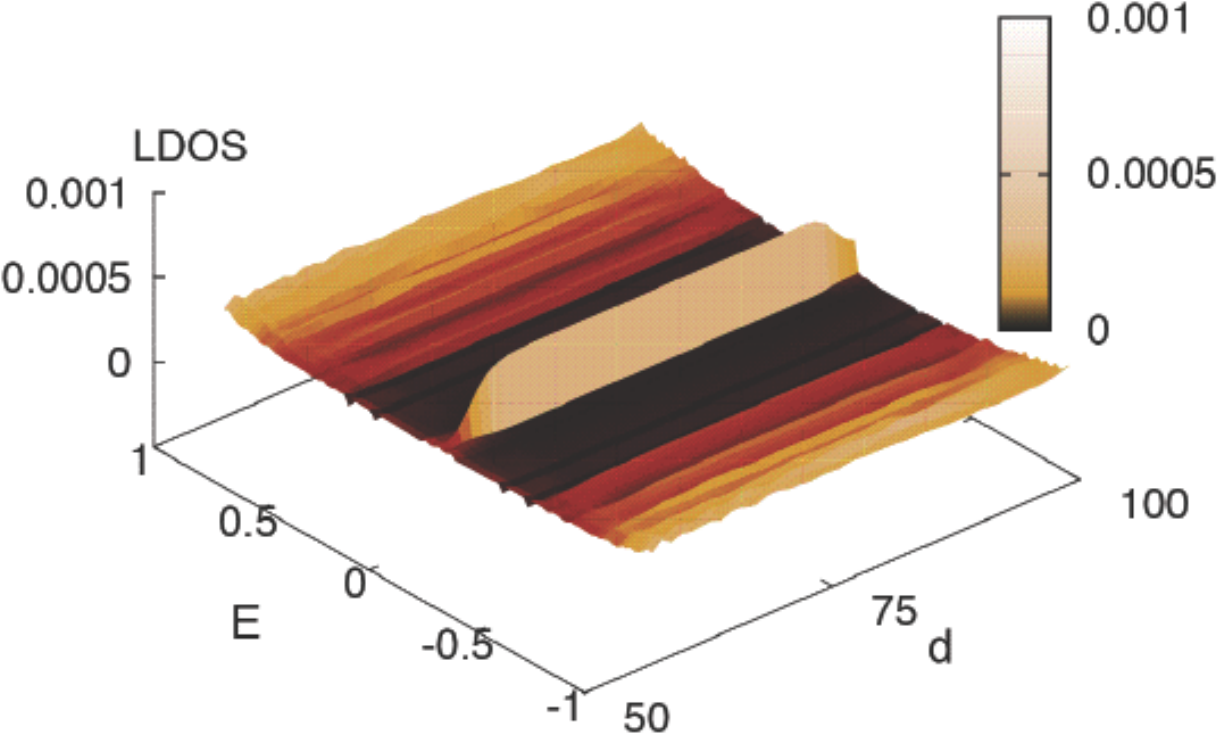}
 \caption{(color online) The LDOS averaged over the six lattice sites of each hexagon along the line (1) of Fig. (\ref{lattice_fur}) is shown.
The combined effect of the mixed edge is that each atom is either
B (A') type or A (B') type which reduces the LDOS
at E=0. So when we sum the LDOS of each lattice sites of a hexagon
close to the left end of line (1), we do not see the LDOS peak at
$E=0$. On the right end of the line where the zigzag and armchair edges cross,
the LDOS peak close to the zigzag edge is absent  due to the presence of the
armchair edge. In the middle of the  line (1) we see the effect of
the proximal zigzag edge. For this calculation we use $\phi=\phi_0/50$.}
 \label{fura}
\end{figure}

In Fig. (\ref{fura}) we show the LDOS averaged over six lattice
sites of the hexagonal cells which lie on the line (1) as shown in Fig.
(\ref{lattice_fur}).
Along this line there is a tilted zigzag edge on the left, a horizontal zigzag edge parallel and close to it and an armchair edge
 on the right side. At the left side, each atom sees the effect of both the horizontal and the tilted zigzag
 edge.  For example, let us have a close look at the cell denoted numerically by ``1'' in Fig.
(\ref{lattice_fur}); The atom at the top left corner of this cell is of type B for the horizontal zigzag edge and of type A'
 for the tilted zigzag edge. Because of the horizontal zigzag edge the LDOS at
 this sublattice should have zero magnitude at $E=0$, but from the reference of the tilted edge,
  it has to have an enhanced LDOS at $E=0$.
  The same logic applies to the other sublattices of this cell.
 Because of this competition there will be a destructive interference and LDOS
  will be very small at $E=0$.
   Now if we go towards the right of the line (1), the effect of the tilted zigzag edge should be reduced but the effect
   of the proximal horizontal zigzag edge should still be present. Therefore we see a gradual increase in the LDOS at $E=0$.
     At the far right, we have a proximal horizontal zigzag edge and a vertical armchair edge. For armchair edge
     the LDOS at $E=0$ is small for both A and B sublattices. On the other hand the zigzag edge has a high LDOS at $E=0$ at site A close to the edge.  Because of this competition the LDOS over the hexagonal cell is reduced at the far right end. It is not zero because of the presence of the proximal zigzag edge.
\begin{figure}
  \includegraphics[width=8.5cm, height=5.0cm]{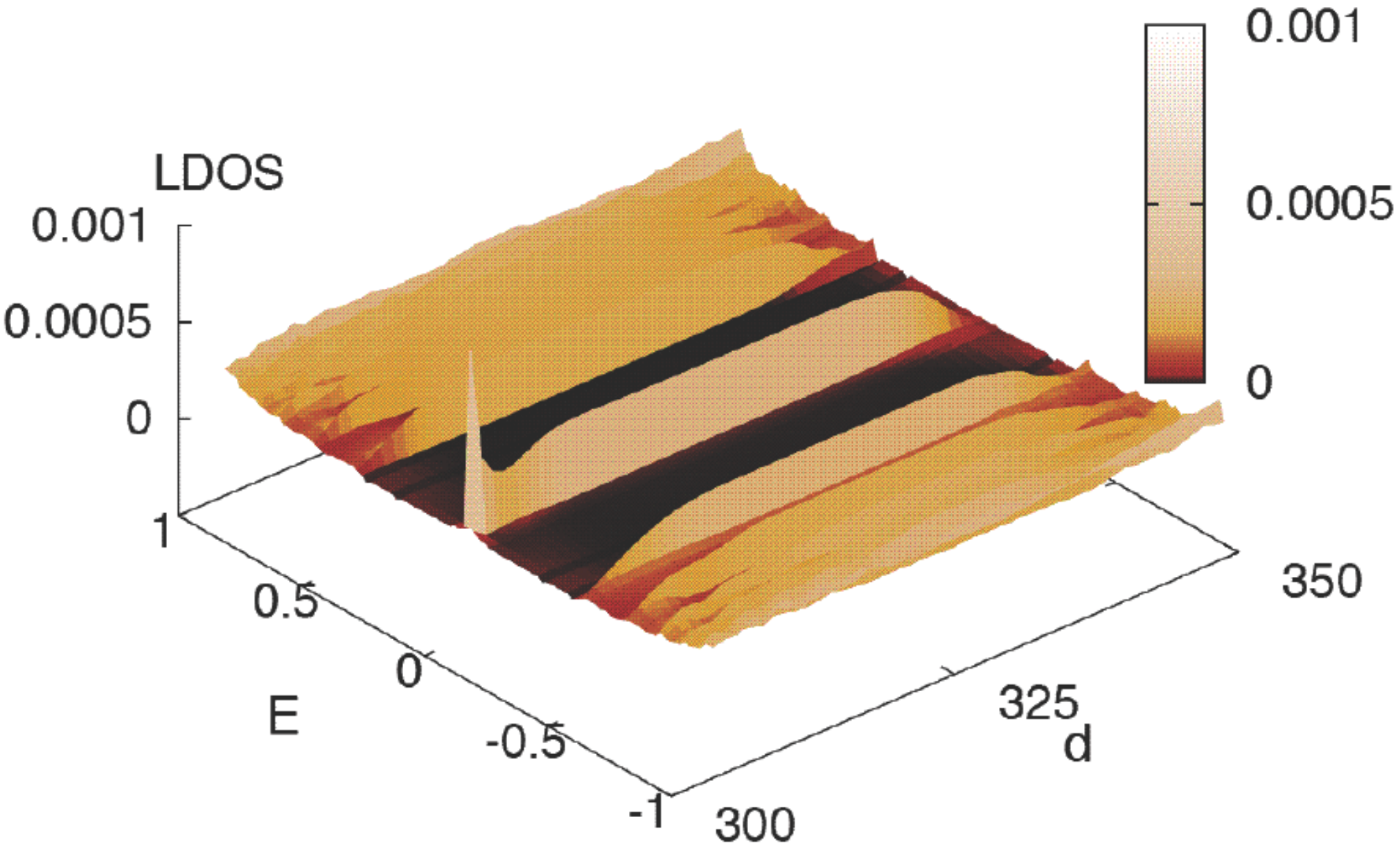}
 \caption{(color online)  The LDOS summed over each lattice sites of the hexagonal cell along the line (2) of Fig. (\ref{lattice_fur}).
  At the far left end of the line there is an enhanced LDOS and at the far right end the LDOS is reduced at $E=0$.
  The tilted zigzag edge at the left end of line (2) is far from the horizontal zigzag edges, and the vertical armchair edges
  leaving the effect of the tilted zigzag edge unaffected. This leads to the appearance of the enhanced LDOS at the left end.
  The LDOS at the far right end is affected by the presence of the armchair edge hence we get almost zero LDOS at $E=0$ at
  this end. In the middle of the line the LDOS shows the behavior of the bulk graphene.The magnet field is $\phi=\phi_0/50$.}
 \label{furb}
\end{figure}

We also calculate the LDOS over the hexagonal cell along the line
(2) of the Fig. (\ref{lattice_fur}). Along this line, there is a tilted zigzag edge on left side, two horizontal zigzag edges
(which are quite far) and an armchair edge on the right end of
the line. The result is shown in Fig. (\ref{furb}). In this case,
the two horizontal zigzag edges are out of the picture. The
appearance of the enhanced LDOS at $E=0$ for smaller 'd' signals the presence of the  tilted zigzag
edge close to the corresponding hexagonal cell. The amplitude of this  LDOS peak decays as we move inside the bulk along the line (2). It is expected because the
effect of the zigzag edge decays away from the edge. Far from the tilted zigzag edge the bulk behavior
characterized by the finite LDOS at $E=0$ is restored. At the far right
end of the line (2) the LDOS at $E=0$ starts to gradually reduce from the bulk value and becomes zero because of the presence of the
armchair edge.

\begin{figure}
  \includegraphics[width=8.5cm, height=5.0cm]{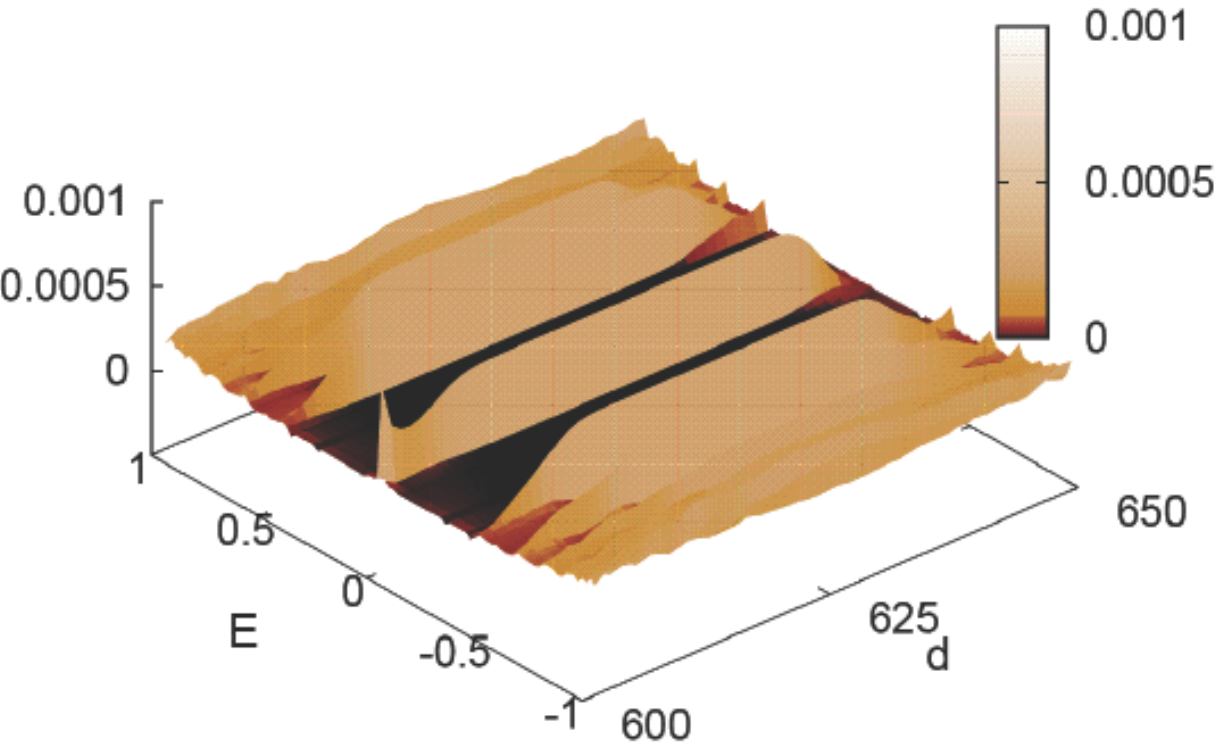}
 \caption{(color online) The LDOS summed over each lattice site of the hexagonal cell along the line (3) of Fig. (\ref{lattice_fur}).
 At the left side of the lattice we see an enhanced LDOS at $E=0$ corresponding to the presence of the tilted zigzag edge and
a reduced LDOS at the other side corresponding to the presence of the armchair edge.
 If we compare the LDOS at the left end of line (2) and (3), we see a reduced amplitude of the LDOS at $E=0$ at the left end of the line
 (3), which is due to the proximity of the tilted edge to the armchair edge.The magnet field is $\phi=\phi_0/50$.} \label{furc}
\end{figure}

We also calculate the LDOS in the scenario when the tilted zigzag edge is close to the armchair edge, e.g. along the line (3) in Fig. (\ref{lattice_fur}) .
 To the left  of this line we have tilted zigzag edge in the proximity of the armchair edge. The two horizontal zigzag edges
  are far from it so they are out of the picture. There is an armchair edge at the far right of the line. The result is shown in Fig. (\ref{furc}).
  The LDOS peak at $E=0$ at the far left of the line is due to the presence of the tilted zigzag edge. But
   the amplitude is reduced compared to that of the hexagonal cell which is at the left of the line (2) because of the presence of the proximal armchair edge.
   At the far right of the line (3) the LDOS is suppressed due to the presence of the armchair edge.
   In the middle of the line we see the bulk behavior.

\begin{figure}
  \includegraphics[width=8.5cm, height=5.0cm]{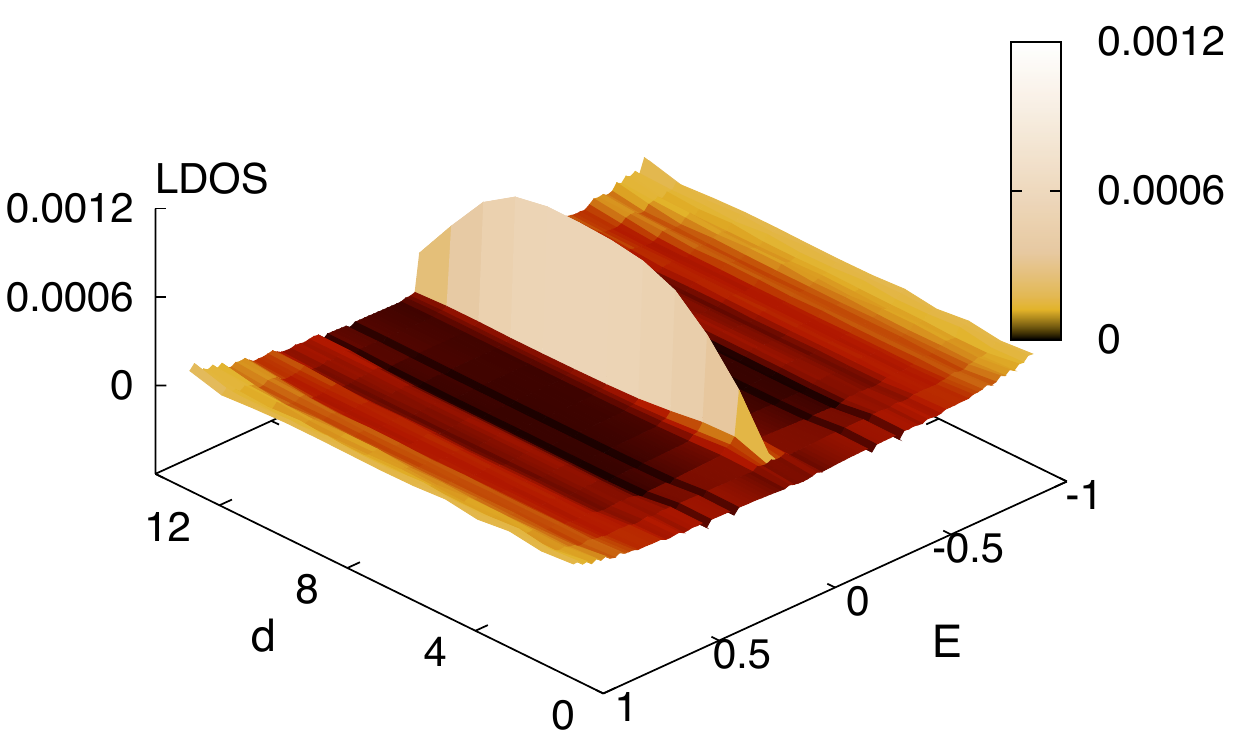}
 \caption{\label{furred}(color online)  The LDOS in the middle of the new zigzag edge along tilted zigzag edge (greenline) of the Fig. (\ref{lattice_fur}).
 The hexagonal cells denoted by $d=1,2,3,......,14$ are the leftmost unit cell in the titled edge. The LDOS at $E=0$ first decreases due to the destructive interference caused by the presence of the horizontal and tilted zigzag edge. As we go inside
 the lattice along Y-axis the role of the horizontal zigzag edge decreases and the role of the titled zigzag edge increases leading
  to the enhanced LDOS at $E=0$, where, in the absence of the titled edge the LDOS would behave as that of the bulk graphene.
  As we move close to the left most armchair edge the amplitude of the LDOS again decreases. The magnet field is $\phi=\phi_0/50$.}
\end{figure}

To sum up the effects of the mixed edge on the LDOS, we calculate
the LDOS along the tilted edge which is represented by the zigzag green line in Fig. (\ref{lattice_fur}). The result is shown in
Fig. (\ref{furred}).
We label the hexagonal cells along the titled zigzag edge by a set of numbers $d=1,2,3,......,14$. We see that when the hexagonal cell is close to the intersection of the horizontal
  and the tilted zigzag edge (close to the top-right) the LDOS at $E=0$ is very small. As we move
along this tilted edge the LDOS at $E=0$ increases. This is expected since we have a tilted zigzag
edge which is far from the horizontal zigzag edge and also far
from the vertical armchair edge.
Close to the bottom-left end of the tilted zigzag edge
the weight of the LDOS at $E=0$ decreases but remains finite.
The reduction is due to the presence of the proximal armchair edge.

The edge states of a mixed edge graphene lattice can be studied analytically using appropriate boundary conditions. Such boundary conditions
have been derived by Akhmerov and Beenakker. \cite{akhmerov} The calculation of the dispersion relation for such mixed edges shows that the surface states of the perfect zigzag edge become dispersive. In terms of the density of states, it is shown
that the density of the edge states is maximum for a perfect zigzag edge and decreases continuously when the mixed edge has less and less zigzag bonds and more and more armchair bonds. Our result is consistent with this calculation since 
we have shown that the height of the local density of states peak at $E=0$ near the edge is either reduced or completely washed out when the lattice has mixed edges.   

 \section{Conclusion}
 Most of the previous studies of the edge states in graphene have been done with considerations
that the graphene lattice has only one type of edge, either armchair or zigzag, and that the hopping energy of electrons is isotropic. In reality the
hopping can be anisotropic and the edges can be mixed. In this paper we study the edge states considering  a lattice which has both anisotropic electron hopping
and mixed edges. Our focus was on the behavior of the LDOS in the vicinity of the edges.

We show that the band structure of the armchair edge graphene changes
qualitatively when hopping becomes anisotropic. It
develops zero energy states. The local density of states also
changes. It now gives rise to the enhanced LDOS at $E=0$ which is
similar to what zigzag edge does whether there is isotopic hopping
or the anisotropic hopping. The amplitude of the enhanced LDOS in
the strained armchair edge graphene depends on the degree of the
anisotropy.

In the mixed edge structure we assumed to have a tilted zigzag edge which meets with the
 horizontal zigzag edge at one side and with the armchair edge at the other side.
  In this situation,  we show that a) near the crossing of the horzontal and slanted zigzag edges the enhanced LDOS at E=0 is completely suppressed,
b) near the crossing of the armchair and slanted zigzag edges the LDOS at  E=0  is smeared out, and
c) near the crossing of the horizontal zigzag and vertical arm chair edges, the LDOS at E=0 at the armchair edge side is enhanced.
   So, depending upon the structure of the lattice and the position where the LDOS is going to be probed,
   the surface state of the zigzag edge can have weight between zero to some maximum values.
   Our results are important to interpret data of the LDOS measurements in the STM experiments.

This work was carried out under the auspices of the National
Nuclear Security Administration of the U.S. Department of Energy
at Los Alamos National Laboratory under Contract No.
DE-AC52-06NA25396. Z.X.H thanks the Ministry of Education of China
for support to visit the NHMFL. K. Y. is supported by National Science Foundation Grant No.
DMR-0704133.

\end{document}